\documentclass[doublecol]{epl2}

\title{Stochastic jump processes for non-Markovian quantum dynamics}

\author{H.-P. Breuer\inst{1}\thanks{E-mail: \email{breuer@physik.uni-freiburg.de}}
\and J. Piilo\inst{2}\thanks{E-mail: \email{jyrki.piilo@utu.fi}}}
\shortauthor{H.-P. Breuer and J. Piilo}

\institute{
  \inst{1} Physikalisches Institut, Universit\"at Freiburg - Hermann-Herder-Strasse 3,
           D-79104 Freiburg, Germany\\
  \inst{2} Department of Physics and Astronomy, University of Turku - FI-20014
           Turun yliopisto, Finland}


\abstract{It is shown that non-Markovian master equations for an
open system which are local in time can be unravelled through a
piecewise deterministic quantum jump process in its Hilbert space.
We derive a stochastic Schr\"odinger equation that reveals how
non-Markovian effects are manifested in statistical correlations
between different realizations of the process. Moreover, we
demonstrate that possible violations of the positivity of
approximate master equations are closely connected to
singularities of the stochastic Schr\"odinger equation, which
could lead to important insights into the structural
characterization of positive non-Markovian equations of motion.}

\begin{document}

\maketitle

Relaxation and decoherence phenomena in open quantum systems
\cite{Breuer2002} can often be modelled with sufficient accuracy
by a quantum Markov processes in which the open system's density
matrix is governed by a relatively simple quantum Markovian master
equation with Lindblad structure \cite{GORINI,Lindblad}. However,
non-Markovian quantum systems featuring strong memory effects play
an increasingly important role in many fields of physics such as
quantum optics~\cite{Gardiner96a}, solid state physics~\cite{SS},
and quantum information science~\cite{QIP}. Further applications
include non-Markovian extensions of quantum process tomography,
quantum control \cite{CONTROL}, and quantum transport
\cite{TRANSPORT}.

The non-Markovian quantum dynamics of open systems is
characterized by pronounced memory effects, finite revival times
and non-exponential behavior of damping and decoherence, resulting
from long-range correlation functions and from the dynamical
relevance of large correlations and entanglement in the initial
state. As a consequence the theoretical treatment of non-Markovian
quantum dynamics is generally extremely demanding, both from the
analytical and from the computational point of view~\cite{HPB08}.
Even if one is able to derive an appropriate non-Markovian master
equation or some other mathematical formulation of the dynamics,
the numerical simulation of such processes turns out to be a very
difficult and time-consuming task, especially for high-dimensional
Hilbert spaces.

From classical physics it is known that Monte Carlo techniques
provide efficient tools for the numerical simulation of complex
systems. This fact was one of the motivations to introduce the
Monte Carlo wave function technique
\cite{DCM1992,Dum1992,Carmichael1993} which provides efficient
quantum simulation techniques in the regime of Markovian dynamics.
Several generalizations of the Monte Carlo approach to
non-Markovian dynamics have been developed which are based on
suitable extensions of the underlying reduced system's Hilbert
space \cite{IMAMOGLU,Garraway1997,Breuer99a,Breuer2004}.

Recently, an efficient alternative simulation algorithm for the
treatment of non-Markovian open system dynamics has been proposed
\cite{Piilo2007} that does not require any extension of the state
space. The purpose of the present paper is to develop a
mathematical formulation of this algorithm in terms of a
stochastic Schr\"odinger equation (SSE) in the open system's
Hilbert space. We demonstrate that this formulation gives rise to
a new type of piecewise deterministic quantum jumps process (PDP).

Quantum master equations are often derived from an underlying
microscopic theory by employing some approximation scheme. An
appropriate scheme is the time-convolutionless (TCL) projection
operator technique which leads to a time-local first-order
differential equation for the density matrix
\cite{KUBO63,ROYER72,SHIBATA2}. It will be shown that TCL
master equations allow a stochastic unravelling of the form
developed here. Generally, the use of a certain approximation
technique may lead to violations of the positivity of the master
equation. We demonstrate that positivity violations are closely
linked to singularities of the SSE at which the stochastic process
breaks down. Hence, a great advantage of the present stochastic
formulation consists in the fact that it naturally prevents the
generation of unphysical solutions and that it could thus lead to
important insights into the structural characterization of
positive non-Markovian evolution equations.

The most general structure of the TCL master equation is given by
\footnote{This statement is a direct consequence of Lemma 2.3 of
Ref. \cite{GORINI}; for various alternative forms see
\cite{Breuer2002,Breuer2004}.}
\begin{equation} \label{MASTER-EQ}
 \frac{d}{dt}\rho(t) = {\mathcal{L}}_t\rho(t)
 = -i[H(t),\rho(t)] + {\mathcal{D}}_t\rho(t),
\end{equation}
where
 \begin{eqnarray*}
 {\mathcal{D}}_t\rho = \sum_m \Delta_m(t)
 \left[ C_m(t)\rho C_m^{\dagger}(t)
 - \frac{1}{2} \left\{C_m^{\dagger}(t)C_m(t),\rho \right\}\right].
\end{eqnarray*}
The time-dependent generator ${\mathcal{L}}_t$ consists of a
commutator term describing the unitary part of the evolution and a
dissipator ${\mathcal{D}}_t$. The latter involves a summation over
the various decay channels labelled by $m$ with corresponding
time-dependent decay rates $\Delta_m(t)$ and arbitrary
time-dependent system operators $C_m(t)$.

In the simplest case the rates $\Delta_m$ as well as the
Hamiltonian $H$ and the operators $C_m$ are assumed to be
time-independent. Equation (\ref{MASTER-EQ}) then represents a
master equation in Lindblad form \cite{GORINI,Lindblad} which
generates a semigroup of completely positive dynamical maps known
as quantum Markov process. For arbitrary time-dependent operators
$H(t)$ and $C_m(t)$, and for $\Delta_m(t) \geq 0$ the generator of
the master equation (\ref{MASTER-EQ}) is still in Lindblad form
for each fixed time $t$ and leads to a two-parameter family of
completely positive dynamical transformations \cite{SPOHN} which
may be referred to as time-dependent quantum Markov process
\cite{EISI}. An entirely different situation emerges if one or
several of the $\Delta_m(t)$ become temporarily negative which
expresses the presence of strong memory effects in the reduced
system dynamics. The process is then said to be non-Markovian. Of
course, the physical interpretation of the master equation
requires that it preserves the positivity of the density matrix
$\rho$. The formulation of general mathematical and physical
conditions that guarantee the preservation of positivity is,
however, an unsolved problem of central importance in the field of
non-Markovian quantum dynamics.
We emphasize that the emergence of temporarily negative
$\Delta_m(t)$ in the master equation is a natural phenomenon in
the non-Markovian regime which does in general not imply that the
complete positivity of the corresponding quantum dynamical map is
violated. An example is discussed in Ref.~\cite{Breuer2002} where
the exact non-Markovian master equation of an analytically
solvable model is constructed.

The fundamental difference between Markovian and time-dependent
Markovian processes on the one hand and non-Markovian processes on
the other hand can also be seen very clearly if one attempts to
apply the standard stochastic formulations to the master equation
(\ref{MASTER-EQ}). For both a Markovian and a time-dependent
Markovian dynamics the standard unravelling through a stochastic
quantum jump process can indeed be applied. This means that in
both cases one can formulate an appropriate PDP for the reduced
system's state vector $|\psi(t)\rangle$ in such a way that the
expectation value
\begin{equation} \label{EXPEC}
 \rho(t) = {\mathrm{E}}[|\psi(t)\rangle\langle\psi(t)|]
 = \int d\psi \, P[|\psi\rangle,t] \, |\psi\rangle\langle\psi|
\end{equation}
satisfies the master equation (\ref{MASTER-EQ}). Here, we have
expressed the expectation value ${\mathrm{E}}$ in terms of an
integration over the Hilbert space of states of the open quantum
system 
with the unitarily invariant volume element
$d\psi \equiv D\psi D\psi^*$,
and introduced the corresponding
probability density functional $P[|\psi\rangle,t]$ which is
defined as the probability density of finding at time $t$ the
state vector $|\psi\rangle$ \cite{Breuer2002}. However, the
essential feature of a non-Markovian dynamics is the temporary
appearance of negative decay rates. The use of the standard
unravellings unavoidably leads in this case to negative jump
probabilities, which clearly indicates the decisive difference
between Markovian and non-Markovian quantum processes.

To account for the sign of the decay rates we decompose
$\Delta_m(t)$ into a positive and a negative part defined by
$\Delta_m^{\pm}(t)=\frac{1}{2}\big[|\Delta_m(t)|\pm\Delta_m(t)\big]$.
The master equation (\ref{MASTER-EQ}) can then be written in the
form
\begin{eqnarray} \label{MASTER-EQ2}
 \frac{d}{dt}\rho &=& -i[H(t),\rho] \\
 &+& \sum_k\Delta_k^+(t) \left[ C_k(t)\rho C_k^{\dagger}(t)
 - \frac{1}{2} \left\{C_k^{\dagger}(t)C_k(t),\rho \right\}
 \right] \nonumber \\
 &-& \sum_l \Delta_l^-(t)\left[ C_l(t)\rho C_l^{\dagger}(t)
 - \frac{1}{2} \left\{C_l^{\dagger}(t)C_l(t),\rho \right\}
 \right]. \nonumber
\end{eqnarray}
In order to better distinguish the positive and the negative
channels we label the former by an index $k$ and the latter by an
index $l$. Note that $\Delta_k^+(t)\geq 0$ and $\Delta_l^-(t)\geq
0$ and that for Markovian or time-dependent Markovian processes we
have $\Delta_l^-(t)=0$.

We can now formulate the central result of this paper. Namely, the
stochastic Schr\"odinger equation given by
 \begin{eqnarray} \label{SSE}
 d|\psi(t)\rangle &=& -i G(t) |\psi(t)\rangle dt  \nonumber \\
 &+&\sum_k\left[\frac{C_k(t)|\psi(t)\rangle}{||C_k(t)|\psi(t)\rangle||}
 - |\psi(t)\rangle \right] dN_k^+(t)  \nonumber \\
 &+&\sum_l \int d\psi' \left[ |\psi'\rangle - |\psi(t)\rangle \right]
 dN_{l,\psi'}^-(t)
\end{eqnarray}
yields an unravelling of the master equation (\ref{MASTER-EQ2})
through a non-Markovian piecewise deterministic process. The first
term on the r.h.s. of Eq.~(\ref{SSE}) represents the normalized
deterministic drift of the process which is generated by
\begin{eqnarray*}
 G(t) &=& H(t) - \frac{i}{2} \sum_m \Delta_m(t) \\
 && \times \Big[ C_m^{\dagger}(t)C_m(t)
 - \langle\psi(t)|C_m^{\dagger}(t)C_m(t)|\psi(t)\rangle \Big].
\end{eqnarray*}
The instantaneous and random quantum jumps are described by the
second and the third line of Eq.~(\ref{SSE}). The quantities
$dN_k^+(t)$ and $dN_{l,\psi'}^-(t)$ are random Poisson increments
satisfying the relations
\begin{eqnarray} \label{PROPS}
 dN_k^+(t)  dN_{k'}^+(t) &=& \delta_{kk'}dN_k^+(t), \nonumber \\
 dN_{l,\psi'}^-(t) dN_{l',\psi''}^-(t)
 &=&  \delta_{ll'} \delta\big(|\psi'\rangle-|\psi''\rangle\big)
 dN_{l,\psi'}^-(t), \nonumber \\
 dN_k^+(t) d N_{l,\psi'}^-(t)& =& 0,
\end{eqnarray}
and having expectation values
\begin{eqnarray}
 {\mathrm{E}}[dN_k^+(t)] &=&
 \Delta_k^+(t)\langle\psi(t)|C_k^{\dagger}(t)C_k(t)|\psi(t)\rangle dt,
 \label{EXPEC1} \\
 {\mathrm{E}}[dN_{l,\psi'}^-(t)] &=& \Delta_l^-(t)
 \frac{P\left[|\psi'\rangle, t\right]}{P\left[|\psi\rangle, t\right]}
 \langle\psi'|C_l^{\dagger}(t)C_l(t)|\psi'\rangle \nonumber \\
 && \times \; \delta\left( |\psi(t)\rangle -
 \frac{C_l(t) |\psi'\rangle}{||C_l(t) |\psi'\rangle||}\right) dt.
 \label{EXPEC2}
\end{eqnarray}
Here, the delta functional on Hilbert space is defined by $\int
d\psi \delta(|\psi\rangle-|\psi_0\rangle) F[|\psi\rangle] =
F[|\psi_0\rangle]$, where $F[|\psi\rangle]$ is an arbitrary smooth
functional.

The physical meaning of the properties in (\ref{PROPS}) is that
there cannot be two or more jumps simultaneously in a given
realization of the process and in a given moment of time. Suppose
first that the dynamics is Markovian or time-dependent Markovian.
We then have $dN_{l,\psi'}^-(t)=0$ and the SSE (\ref{SSE}) reduces
to the stochastic differential equation of the standard PDP
unravelling. According to the second line of Eq.~(\ref{SSE}) the
quantum jumps are represented by an instantaneous change of the
state vector,
\begin{eqnarray*}
 |\psi(t)\rangle \longrightarrow
 \frac{C_k(t)|\psi(t)\rangle}{||C_k(t)|\psi(t)\rangle||},
\end{eqnarray*}
and by virtue of Eq.~(\ref{EXPEC1}) this jump occurs at the rate
\begin{equation} \label{RATE1}
 \Gamma_+ = \Delta_k^+(t)\langle\psi(t)|C_k^{\dagger}(t)C_k(t)|\psi(t)\rangle.
\end{equation}

The term in the third line of the SSE (\ref{SSE}) describes the
negative channels which are crucial for the unravelling of
non-Markovian dynamics. The corresponding jumps are given by
instantaneous transitions from the actual state $|\psi(t)\rangle$
to some state $|\psi'\rangle$. To account for all possible target
states of the negative channel jumps we perform in this term an
integration over $|\psi'\rangle$. According to the delta
functional in Eq.~(\ref{EXPEC2}) the target state $|\psi'\rangle$
of the possible jumps is related to the source state
$|\psi\rangle$ by
$|\psi\rangle=C_l|\psi'\rangle/||C_l|\psi'\rangle||$. Hence,
negative jumps correspond to a reversal of certain positive jumps,
obtained by interchanging the role of source and target state. The
quantity $dN_{l,\psi'}^-(t)$ is the Poisson increment for the
negative jumps via channel $l$. From Eq.~(\ref{EXPEC2}) we infer
that the state vector $|\psi\rangle$ can perform a jump to a state
vector in some volume element $d\psi'$ of Hilbert space around
$|\psi'\rangle$ with the rate (for simplicity we omit the time
arguments)
\begin{eqnarray*}
 \Gamma_- =
 \Delta_l^- \frac{P\left[|\psi'\rangle\right]d\psi'}{P\left[|\psi\rangle\right]d\psi}
 \langle\psi'|C_l^{\dagger}C_l|\psi'\rangle
 \delta\left( |\psi\rangle - \frac{C_l |\psi'\rangle}{||C_l |\psi'\rangle||}\right)
 d\psi.
\end{eqnarray*}
In an ensemble of realizations of the process the quantity
$P[|\psi'\rangle]d\psi'/P[|\psi\rangle]d\psi$ can be interpreted
as $N'/N$, where $N'$ is the number of realizations in volume
element $d\psi'$ and $N$ is the number of realizations in element
$d\psi$. Then we can identify $\delta \left( |\psi\rangle - C_l
|\psi'\rangle / ||C_l|\psi'\rangle||\right) d\psi = 1$. Hence, the
negative channel jumps from $|\psi\rangle$ to $|\psi'\rangle$
occur at the rate
\begin{equation} \label{RATE2}
 \Gamma_- = \Delta_l^- \frac{N'}{N}
 \langle\psi'|C_l^{\dagger}C_l|\psi'\rangle.
\end{equation}
The comparison with the rate (\ref{RATE1}) for positive jumps
shows two crucial differences. First, the rates for the positive
jumps is proportional to the expectation value of
$C_k^{\dagger}C_k$ in the source state, while the rates for the
negative jumps is proportional to the expectation value of
$C_l^{\dagger}C_l$ in the target state. Hence, again the role of
source and target state have been interchanged. Second, the
negative jump rates carry an additional factor of $N'/N$, the
ratio of the number of ensemble members in the target state to the
number of members in the source state. Note that due to the
presence of this factor the SSE (\ref{SSE}) is not a stochastic
differential equation in the usual sense because the expectation
values of the random increments (\ref{EXPEC2}) depend explicitly
on the full probability density. 
To determine these
increments at a certain time $t$ one has to know the probability
density $P[|\psi\rangle,t]$. Within a numerical simulation this is
achieved by propagating simultaneously an ensemble of realization
from which $P[|\psi\rangle,t]$ can then be estimated
self-consistently.
As an important consequence and as a result of
the non-Markovian character of the dynamics we thus find certain
correlations between different realizations of the process.

It may seem at first sight that the correlations between
the realizations require that a huge number of realizations of the
process has to be generated simultaneously in order to obtain a
sufficiently accurate estimate for the probability density.
However, when the realizations of the process are generated on a
computer there is no need to have $N_i$ identical copies of the
state $|\psi_i\rangle$ to obtain $P\left[|\psi_i\rangle\right]$.
It is sufficient to have only a single copy of $|\psi_i\rangle$
and to keep track of the corresponding integer number $N_i$. This
allows to optimize the numerical implementation of the process and
to perform simulations in an efficient way~\cite{Piilo2007}.

To prove that the expectation value (\ref{EXPEC}) for the process
obtained from the SSE (\ref{SSE}) indeed satisfies the master
equation (\ref{MASTER-EQ2}) we start from
\begin{eqnarray*}
 d(|\psi\rangle\langle\psi|) =
 |d\psi\rangle\langle\psi| + |\psi\rangle\langle d\psi|
 + |d\psi\rangle\langle d\psi|.
\end{eqnarray*}
Taking the expectation value of this relation, expressing the
increments $|d\psi\rangle$ through the SSE (\ref{SSE}), and using
the properties (\ref{PROPS}) we find
\begin{eqnarray*}
 d\rho &=& -i[H,\rho]dt - \sum_m\frac{\Delta_m}{2} \left\{C_m^{\dagger}C_m,\rho \right\} dt
 \nonumber \\
  &+& \sum_m\Delta_m \; {\mathrm{E}}\left[ \langle\psi|C_m^{\dagger}C_m|\psi\rangle
 |\psi\rangle\langle\psi|\right] dt \nonumber \\
  &+& {\mathrm{E}}\Bigg[ \sum_k\left( \frac{C_k|\psi\rangle\langle\psi|C_k^{\dagger}}
 {||C_k|\psi\rangle||^2} - |\psi\rangle\langle\psi|
 \right) dN_k^+ \Bigg] \\
 &+&{\mathrm{E}}\left[ \sum_{l} \int d\psi' \left(
 |\psi'\rangle\langle\psi'| - |\psi\rangle\langle\psi| \right)
 dN_{l,\psi'}^- \right].
 \end{eqnarray*}
Using here the expectation values of the increments from
Eq.~(\ref{EXPEC2}) one immediately obtains the required master
equation (\ref{MASTER-EQ2}). Hence, we have proven the validity of
SSE (\ref{SSE}) which is the central result of the paper.
It is worth mentioning that while there exists
stochastic Schr\"odinger equations of diffusion type for
non-Markovian systems~\cite{Strunz1,Strunz2}, to the best of our
knowledge our SSE is the first representation through a stochastic
quantum jump process in the reduced system's Hilbert space.

It is important to note that the expectation value for
$dN_{l,\psi'}^-(t)$ in Eq.~(\ref{EXPEC2}) is not well defined when
the denominator becomes equal to zero, i.~e. $P\left[|\psi\rangle,
t\right]=0$, where $|\psi\rangle =C_l |\psi'\rangle  / ||C_l
|\psi'\rangle||$, or alternatively $N=0$ in Eq.~(\ref{RATE2}). The
stochastic process breaks down at this point since there exists an
open negative channel but there are no realizations which are in
the source state of the corresponding jump. The formulation of
general conditions on the structure of the master equation
(\ref{MASTER-EQ}) that ensure the absence of such singularities in
the corresponding SSE (\ref{SSE}) seems to be a difficult problem.
However, it is quite easy to demonstrate that a breakdown of the
process necessarily takes place if the master equation violates
positivity. In fact, within the stochastic formulation developed
here the density matrix $\rho(t)$ of the open system is given by
the expectation value (\ref{EXPEC}) which represents, by the very
construction, a positive matrix. Therefore, if the master equation
violates positivity at some point of time the stochastic dynamics
must necessarily cease to exist. The present method thus signals
the point of violation of positivity of the density matrix.

To prove this statement we denote the state space, i.~e., the set
of all density matrices of the open system by ${\mathcal{S}}$. Let
us assume that the master equation (\ref{MASTER-EQ}) violates
positivity. Hence, there is an initial state $\rho(0)$ and a
corresponding solution $\rho(t)$ of the master equation which
leaves the state space ${\mathcal{S}}$ after some point of time
$t=t_0$. At this point $\rho(t_0)=\rho_0$ reaches the boundary of
${\mathcal{S}}$. Let
$\lambda(t)=\langle\varphi(t)|\rho(t)|\varphi(t)\rangle$ be the
lowest eigenvalue of $\rho(t)$ with corresponding eigenvector
$|\varphi(t)\rangle$. Lying on the boundary, $\rho_0$ must have at
least one zero eigenvalue with corresponding eigenvector
$|\varphi_0\rangle=|\varphi(t_0)\rangle$, i.~e., we have
$\lambda(t_0)=\langle\varphi_0|\rho_0|\varphi_0\rangle=0$. Hence,
an appropriate condition implying the violation of positivity is
given by the inequality $\dot{\lambda}(t_0)<0$~\footnote{It is
assumed here that $\dot{\lambda}(t_0)$ does not vanish, which is
obviously the generic case.}. The Hellman-Feynman theorem yields
\begin{eqnarray*}
 \dot{\lambda}(t_0) = \langle\varphi_0|\dot{\rho}(t_0)|\varphi_0\rangle
 = \langle\varphi_0|{\mathcal{L}}_{t_0}\rho_0|\varphi_0\rangle,
\end{eqnarray*}
and we find the following condition for the violation of positivity
\begin{equation} \label{COND}
 \langle\varphi_0|{\mathcal{L}}_{t_0} \rho_0 |\varphi_0\rangle < 0.
\end{equation}

Consider now an ensemble representation of $\rho_0$ that is
generated through the SSE (\ref{SSE}): $\rho_0 = \sum_i p_i
|\psi_i\rangle\langle\psi_i| $ with
$\langle\psi_i|\psi_i\rangle=1$, $p_i > 0$ and $\sum_i p_i=1$. We
then have
\begin{eqnarray*}
 \langle\varphi_0|\rho_0|\varphi_0\rangle = \sum_i p_i
 |\langle\varphi_0|\psi_i\rangle|^2 = 0.
\end{eqnarray*}
It follows that $|\varphi_0\rangle$ is orthogonal to all members
of the ensemble, i.~e., $\langle\varphi_0|\psi_i\rangle=0$.
Evaluating condition (\ref{COND}) one therefore finds
\begin{eqnarray*}
 \langle\varphi_0|{\mathcal{L}}_{t_0} \rho_0 |\varphi_0\rangle
 = \sum_{m,i} p_i \Delta_m(t_0) |\langle\varphi_0|C_m|\psi_i\rangle|^2
 < 0.
\end{eqnarray*}
Hence, there must exist indices $m$ and $i$ such that
$\Delta_m(t_0)<0$ and $\langle\varphi_0|C_m|\psi_i\rangle\neq 0$.
It follows that $C_m|\psi_i\rangle$ has a nonzero component in the
direction of $|\varphi_0\rangle$ and that the state vector
$|\psi\rangle=C_m|\psi_i\rangle/||C_m|\psi_i\rangle||$ does not
belong to the ensemble $\{|\psi_i\rangle\}$. In other words,
$P\left[|\psi\rangle,t_0 \right]=0$. We conclude that the point of
violation of positivity implies the breakdown of the SSE
(\ref{SSE}) because there exists an open channel with negative
rate while the probability of being in the source state of the
corresponding jump vanishes.

Of course, the formal mathematical solution of the master equation
(\ref{MASTER-EQ}) does not halt at the point of time when the
positivity is lost: The dynamics continues to reduce occupation
probability of a given state beyond the zero-point. However, the
evolution given by the SSE (\ref{SSE}) stops at the zero-point
since the number of realizations in a given state cannot, by
construction, have negative values. Thus, the stochastic process
developed here identifies the point of time where the master
equation loses the positivity, preventing excursions to unphysical
solutions. While the stochastic unravelling of the
master equation is in general not unique, we expect that the
connection between a breakdown of the positivity and a singularity
of the SSE holds for all stochastic representation of the form
constructed here.

In conclusion, we have derived a piecewise deterministic process
which describes the dynamics of non-Markovian systems. The
stochastic Schr\"odinger equation constructed reveals the
fundamental mathematical and physical difference between
time-local master equations which appear with positive and with
negative rates. The corresponding Poisson increments have a
distinct structure and the negative rate process clearly shows how
non-Markovian effects are manifested.

Markovian and non-Markovian processes are widely used for the
modelling of dynamical systems in many areas of physics, chemistry
and biophysics. Our results indicate how to treat master equations
with negative rates and memory effects also for classical systems.
In fact, the standard simulation algorithm for a classical
Markovian master equation corresponding to the stochastic wave
function method is known as Gillespie algorithm \cite{Gillespie}.
The method proposed here could therefore lead to the development
of an efficient non-Markovian generalization of the Gillespie
algorithm and thus opens the way to many further studies in the
dynamics of complex system.

\acknowledgments
One of us (HPB) greatfully acknowledges financial support
within a Fellowship of the Hanse-Wissenschaftskolleg,
Delmenhorst.
This work has also been supported by the Academy of
Finland (Project No.~115982) and the Magnus Ehrnrooth Foundation.
We thank K.-A. Suominen, S. Maniscalco, and K.
H\"ark\"onen for stimulating discussions.







\begin{thebibliography}{xx}

\bibitem{Breuer2002}
\Name{Breuer H.-P. \and Petruccione F.}
 \Book{The Theory of Open
Quantum Systems}
\Publ{Oxford University Press, Oxford}
\Year{2007}.

\bibitem{GORINI}
\Name{Gorini V., Kossakowski A. \and Sudarshan E. C. G.}
\REVIEW{J. Math. Phys.}{17}{1976}{821}.

\bibitem{Lindblad}
\Name{Lindblad G.}
Commun. Math. Phys. {\textbf{48}}, 119 (1976).

\bibitem{Gardiner96a}
\Name{Gardiner C. W. \and Zoller P.}
 \Book{Quantum Noise}
\Publ{Springer-Verlag, Berlin, 1999}.

\bibitem{SS}
See, e.g.,
\Name{Lai C. W., Maletinsky P., Badolato A.  \and A. Imamoglu}
\REVIEW{Phys. Rev. Lett.}{96}{2006}{167403}, and references therein.

\bibitem{QIP}
See, e.g.,
\Name{Aharonov D. , Kitaev A. \and Preskill J.}
\REVIEW{Phys. Rev.Lett.} {96}{2006}{050504}.

\bibitem{CONTROL}
\Editor{Mancini S., Man'ko V. I.  \and Wiseman H. M.}
Special issue on quantum control,
\REVIEW{J. Opt. B: Quantum
Semiclass. Opt.} {7} {2005}.


\bibitem{TRANSPORT}
\Editor{Breuer H.-P., Gemmer J., Michel M. \and Schollw\"ock U.}
Quantum transport and relaxation: From foundations to
applications at the nanoscale,
\REVIEW{ Eur. Phys. J. Special Topics}
{151} {2007}.

\bibitem{HPB08}
\Name{Breuer H.-P. \and Vacchini B.}
\REVIEW{Phys. Rev. Lett.}{101}{2008}{140402}.

\bibitem{DCM1992}
\Name{Dalibard J.,  Castin Y.  \and M{\o}lmer K.}
\REVIEW{Phys. Rev. Lett.} {68}{1992}{580}.

\bibitem{Dum1992}
\Name{Dum R., Zoller P. \and Ritsch H.}
\REVIEW{Phys. Rev. A} {45}{1992}{4879}.


\bibitem{Carmichael1993}
\Name{Carmichael H.}
\Book{An Open System Approach to Quantum Optics}
\Publ{Springer-Verlag, Berlin}
\Year{1993}.

\bibitem{IMAMOGLU}
\Name{Imamoglu A.}
\REVIEW{Phys. Rev. A} {50}{1994}{3650}.

\bibitem{Garraway1997}
\Name{Garraway B. M.}
\REVIEW{Phys. Rev. A}{55}{1997}{2290}.

\bibitem{Breuer99a}
\Name{Breuer H.-P., Kappler B. \and Petruccione F.}
\REVIEW{Phys. Rev. A}{59}{1999}{1633}.

\bibitem{Breuer2004}
\Name{Breuer H.-P.}
\REVIEW{Phys. Rev. A} {70}{2004}{012106}.

\bibitem{Piilo2007}
\Name{Piilo J., Maniscalco S., H\"arkonen K. \and Suominen K.-A.}
\REVIEW{Phys. Rev. Lett.} {100}{2008}{180402}.

\bibitem{KUBO63}
\Name{Kubo R.}
\REVIEW{J. Math. Phys.}{4}{1963}{174}.

\bibitem{ROYER72}
\Name{Royer A.}
\REVIEW{Phys. Rev. A}{6}{1972}{1741}.

\bibitem{SHIBATA2}
\Name{Chaturvedi S. \and Shibata F.}
\REVIEW{Z. Phys. B} {35}{1979}{297}.

\bibitem{SPOHN}
\Name{Davies E. B. \and Spohn H.}
\REVIEW{J. Stat. Phys.} {19}{1978}{511}.

\bibitem{EISI}
\Name{Wolf M.M., Eisert J., Cubitt T. S. \and Cirac J. I.}
\REVIEW{Phys. Rev. Lett.}{101}{2008}{150402}.

\bibitem{Strunz1}
\Name{Strunz W. T., Di\`osi L. \and Gisin N.}
\REVIEW{Phys. Rev. Lett.}{82}{1999}{1801}.

\bibitem{Strunz2}
\Name{Di\`osi L., Gisin N. \and Strunz W. T.}
\REVIEW{Phys. Rev. A}{58}{1998}{1699}.


\bibitem{Gillespie}
\Name{Gillespie D. T.}
\REVIEW{J. Phys. Chem.}{81}{1977} 2340.

\end{thebibliography}
\end{document}